\documentclass[preprint,aps,
showpacs
]{revtex4}

\usepackage{epsfig}

\input{epsf}

\newcommand{\be}{\begin{equation}}
\newcommand{\ee}{\end{equation}}
\newcommand{\bea}{\begin{eqnarray}}
\newcommand{\eea}{\end{eqnarray}}
\newcommand{\nn}{\nonumber \\}

\newcommand{\tr}{\mbox{tr}}

\begin{document}

\preprint{Guchi-TP-017}
\date{\today%
}
\title{Compactification in deconstructed gauge theory
with topologically non-trivial link fields}

\author{Yoshinori Cho}
\email{b2669@sty.cc.yamaguchi-u.ac.jp}
\affiliation{Graduate School of Science and Engineering, Yamaguchi University, 
Yoshida, Yamaguchi-shi, Yamaguchi 753-8512, Japan}

\author{Nahomi Kan}
\email{b1834@sty.cc.yamaguchi-u.ac.jp}
\affiliation{Graduate School of Science and Engineering, Yamaguchi University, 
Yoshida, Yamaguchi-shi, Yamaguchi 753-8512, Japan}


\author{Kiyoshi Shiraishi}
\email{shiraish@yamaguchi-u.ac.jp}
\affiliation{Faculty of Science, Yamaguchi University,
Yoshida, Yamaguchi-shi, Yamaguchi 753-8512, Japan}

\begin{abstract}

We investigate the mass spectrum of a scalar field in a world
with latticized and circular continuum space where background fields
takes a topological configuration. We find that the mass spectrum is
related to the characteristic values of Mathieu functions.
The gauge symmetry breaking in a similar spacetime is also discussed.


\end{abstract}



\maketitle


\section{Introduction}

The most exotic approach to realize the unification of fundamental 
forces is based on assuming  higher-dimensional gauge theory.
The field in higher
dimensions brings the corresponding Kaluza-Klein (KK)
spectrum in four
dimensional spacetime provided that the extra space is a compact
manifold~\cite{KK}. 

Recently, there appears a novel scheme to describe higher-dimensional
gauge theory,
which is known as deconstruction~\cite{ACG,HL1,HL2}.
A number of copies of a four-dimensional theory
linked by a new set of fields can be viewed as a single gauge theory.
The resulting theory may be almost equivalent to a 
higher-dimensional theory with discretized, or,
latticized extra dimensions. 

In the continuum spacetime, 
if the compactification involves a topologically
non-trivial configuration of gauge fields, the mass spectrum of charged
fields become radically changed from that of the conventional
compactification~\cite{RSS}. 
The cases with non-trivial field strength on a flat torus have also
been considered~\cite{DS}. 
The mass spectrum affects the low mass
degree of freedom as well as the (Casimir-like) quantum energy density.

It is also known, for higher-dimensional non-Abelian theory, a similar
topological configuration gives rise to a symmetry 
breaking~\cite{Manton,Hosotani}. Such an alternative candidate
to the Higgs mechanism is worth studying in higher-dimensional theories
including string(-inspired) theories.

In deconstructed theories, the link fields can form a topological
configuration as a whole in the presence of another compact dimension.
Such a `hybrid' compactification has been studied~\cite{KS} in the other
context. Of course the original motivation of the deconstruction scheme,
which serves a good UV behavior of the theory, must be ignored in the
present case. On the other hand, however, we can regard the hybrid
compactification as a continuum limit of the deconstructed theory. 


In the present paper, we investigate a $U(1)$ gauge theory with a
latticized circle, assuming another circular continuum dimension
and topologically non-trivial background fields. 
We explicitly show the mass spectrum in the backgroud fields, 
which has a certain limit of the continuum theory. 

In Sec.~\ref{sec:2}, we examine the topologically non-trivial
configuration of the link fields in the compactified spacetime.
The mass spectrum of a charged scalar field in this background field is
studied in Sec.~\ref{sec:3}. 
The mass spectrum of the Yang-Mills field in the same background is
studied in Sec.~\ref{sec:4}, where symmetry breaking in this
situation is shown. 
The final section, Sec.~\ref{sec:f}, is devoted to conclusion.

\section{topologically non-trivial configuration}
\label{sec:2}

We begin with the lagrangian for deconstructing $(D+1)$-dimensional
pure $U(1)$ gauge theory~\cite{HL1}:
\bea
{\cal L}_{V}=\sum_{k=1}^{N}\frac{1}{g^2}
\left[-\frac{1}{4}F_{k}^{\mu\nu}F_{k~\mu\nu}-
(D^{\mu}U_{k})^{\dagger}D_{\mu}U_{k}\right]\, ,
\eea
where $g$ is a gauge coupling,
\be
F_{k}^{\mu\nu}=
i\left[\partial^{\mu}-i\tilde{A}_{k}^{\mu}\, ,
\partial^{\nu}-i\tilde{A}_{k}^{\nu}\right]\, ,
\ee
and
\be
D^{\mu}U_{k}=\partial^{\mu}U_{k}-i\tilde{A}_{k}^{\mu}U_{k}+
iU_{k}\tilde{A}_{k+1}^{\mu}\, .
\ee
The labels of the fields are considered as periodic modulo $N$,
{\it e.g.}, $U_{0}\equiv U_N$, $U_{N+1}\equiv U_1$,
and so on.
Further we assume that all $U_k$ have a common absolute value
$|U_k|=f/\sqrt{2}$.

This theory is invariant under the following gauge transformation:
The transformation of gauge fields is
\be
\tilde{A}_{k}^{\mu}\rightarrow
\tilde{A}_{k}^{\mu}+iW_{k}\partial^{\mu}W_{k}^{\dagger}\, ,
\ee
while the link fields $U_{k}$ are transformed as
\be
U_{k}\rightarrow W_{k}U_{k}W_{k+1}^{\dagger}\, ,
\label{gt}
\ee
where absolute values of $W_{k}$'s are unity.

It is known that when all $U_k$ are assumed to equal $f/\sqrt{2}$,
the mass spectrum of the $D$-dimensional gauge field reads~\cite{HL1}
\be
4f^2\sin^2\left(\frac{\pi p}{N}\right)\qquad\qquad p~{\rm
is~an~integer}\, .
\ee

Now we examine the case with another compact dimension.
We suppose that the $z$-direction is periodical, or 
the following identification is assumed:
\be
z\sim z+2\pi R,
\ee
where $R$ can be considered as a radius of a circle.

If the circular dimension exists,
there are other solutions to the equation of motion for $U_k$
with vanishing $A_k^{\mu}$'s.
Then the part of lagrangian density
$(D^{\mu}U_{k})^{\dagger}D_{\mu}U_{k}$ can be rewritten as
$\partial^{\mu}\chi_{k}\partial_{\mu}\chi_{k}/2$,
when we set $U_k\equiv\exp(i\chi_k/f)$.
The equation of motion is led as $\partial^2\chi_k=0$.

The most general background solution is
$\chi_k/f=nz/(NR)+\varphi$, or
\be
U_{k}=\frac{f}{\sqrt{2}}
\exp\left[ i\left(\frac{nz}{NR}+\varphi\right)\right]\, ,
\ee
where $n$ is an integer and $\varphi$ is independent of $z$.
Possible arbitrary phases are gauged away by transformations
(\ref{gt}), except for a common phase $\varphi$. 
For $n\ne 0$, it is irrelevant to the mass spectrum
because the common phase implies only the translation in
the $z$-direction.
The field corresponding to the common phase,
the existence of
which does not require compact continuum dimensions,
was studied by Hill and Leibovich~\cite{HL1,HL2}.
We adopt only the non-zero $n$ in the present paper.

We show how
$U_{k}=\frac{f}{\sqrt{2}}
\exp\left( i\frac{nz}{NR}\right)$ is taken for
a single-valued function with respect to $z$.
When we choose the transformation~(\ref{gt}) with
\be
W_{k}=\exp\left(-2\pi i\frac{nk}{N}\right)\, ,
\ee
the single-valuedness is satisfied as follows:
\be
U_{k}|_{z=2\pi R}=\frac{f}{\sqrt{2}}
\exp\left( 2\pi i\frac{n}{N}\right)\rightarrow\frac{f}{\sqrt{2}}=
U_{k}|_{z=0}\, .
\ee
The periodicity in the latticized dimension, such as $W_N=W_0$,
is hold when $n$ is an integer. 
$n$ is a topological number as in the case with the magnetic flux
in the two dimensional compact space.
Actually, in the limit of $N\rightarrow\infty$ and $N/f=const.$,
our model corresponds to the model with the constant magnetic
field $B=nf/(NR)$ on the two-torus~\cite{DS}.

\section{scalar field}
\label{sec:3}

The lagrangian for deconstructing $(D+1)$-dimensional
scalar field theory is
\bea
{\cal L}_{\phi}&=&
\sum_{k=1}^{N}
\left[-(D^{\mu}\tilde{\phi}_{k})^{\dagger}D_{\mu}\tilde{\phi}_{k}\right]\nn
&+&f\sum_{k=1}^{N}\left(\sqrt{2}\tilde{\phi}_{k}^{\dagger}U_{k}
\tilde{\phi}_{k+1}+
\sqrt{2}\tilde{\phi}_{k}^{\dagger}U_{k-1}^{\dagger}\tilde{\phi}_{k-1}-
2f\tilde{\phi}_{k}^{\dagger}\tilde{\phi}_{k}\right)\, ,
\eea
where
\be
D^{\mu}\tilde{\phi}_{k}=\partial^{\mu}\tilde{\phi}_{k}-
i\tilde{A}_{k}^{\mu}\tilde{\phi}_{k}\, .
\ee
This lagrangian is invariant under the transformation (\ref{gt})
with
\be
\tilde{\phi}_{k}\rightarrow W_{k}\tilde{\phi}_{k}\, .
\label{gts}
\ee

We investigate the mass spectrum of the scalar field
when the $z$-direction is periodic, $z\sim z+2\pi R$, and
the link fields take the topologically non-trivial form,
\be
U_{k}=\frac{f}{\sqrt{2}}
\exp\left( i\frac{nz}{NR}\right)\, ,
\ee
with an integer $n$.

To obtain the eigenfunctions associated with the spectrum,
we expand the scalar field as
\be
\tilde{\phi}_{k}=\frac{1}{\sqrt{N}}\sum_{p=1}^N\phi_{p}
\exp\left[2\pi i\frac{pk}{N}\right]\, .
\label{ef}
\ee
Then the equation of motion for the charged scalar field
reduces to
\be
-\partial_{\mu}^2\phi_p+2f^2\left[1-\cos\left(\frac{2\pi p}{N}+
\frac{nz}{NR}\right)\right]\phi_p=0\, .
\label{eqsc}
\ee
Now we have to find the eigenfunction of the following value equation
\be
M^2\phi_p=-\partial_{z}^2\phi_p+2f^2\left[1-\cos\left(\frac{2\pi p}{N}+
\frac{nz}{NR}\right)\right]\phi_p\, ,
\ee
to obtain the spectrum of the mass $M$ for scalar
fields in $(D-1)$-dimensional spacetime.

The possible eigenfunctions turn out to be the Mathieu
functions~\cite{GR}. They are given by
\be
\phi_{0,m,p}(z)={\rm ce}_{\frac{2mN}{n}}\left(\frac{nz}{2NR}+\frac{\pi
p}{N}+\frac{\pi}{2},\frac{4N^2f^2R^2}{n^2}\right)\qquad
m=0, 1, 2, \ldots\, ,
\ee
and
\be
\phi_{1,m,p}(z)={\rm se}_{\frac{2mN}{n}}\left(\frac{nz}{2NR}+\frac{\pi
p}{N}+\frac{\pi}{2},\frac{4N^2f^2R^2}{n^2}\right)\qquad
m=1, 2, 3,\ldots\, ,
\ee
up to some normalization constants omitted here.
Their eigenvalues $M^2$ are given by
\be
M^2_{0,m}=\frac{1}{R^2}\left[\frac{n^2}{4N^2}a_{\frac{2mN}{n}}({
{4N^2f^2R^2}/{n^2}})+2(fR)^2\right]\qquad m=0, 1, 2, \ldots\, ,
\ee
and
\be
M^2_{1,m}=\frac{1}{R^2}\left[\frac{n^2}{4N^2}b_{\frac{2mN}{n}}({
{4N^2f^2R^2}/{n^2}})+2(fR)^2\right]\qquad m=1, 2, 3,\ldots\, ,
\ee
respectively, where $a_r(q)$ [$b_r(q)$] is a characteristic value which
yields an even [odd] periodic solution of the Mathieu's equation.%
\footnote{In the limit $q\rightarrow 0$, both $a_r(q)$ and $b_r(q)$
approach to $r^2$.}

\begin{figure}[htb]
\centering
\mbox{\epsfbox{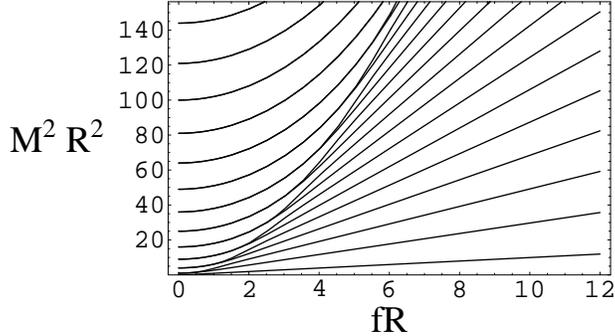}}\\
\bigskip
\caption{%
The mass-squared eigenvalues of the scalar field 
are plotted as functions of
$fR$ for $n=N$.
}
\label{fig1}
\end{figure}

These eigenfunctions do not necessarily satisfy the periodic condition
in the $z$-direction. Therefore the degree of freedom for each eigen
value is determined by the possible combinations in the form~(\ref{ef})
which satisfy the periodic boundary condition.

First we consider the case that the topological number $n$ is a divisor
of $N$, or $n=1$.
In this case, the linear combination
\be
\tilde{\phi}_{a,m,\ell,k}(z)=\sum_{p'=1}^{N/n}
\phi_{a,m,\ell+p'n}(z)
\exp\left[2\pi i\frac{(\ell+p'n)k}{N}\right]\qquad
(a=0, 1)\, ,
\ee
has the correct boundary condition.
To see this, we notice that
$\phi_{a,m,p}(z+2\pi R)=\phi_{a,m,p+n}(z)$ when $N/n$ is an integer.
Then one can find
\be
\tilde{\phi}_{a,m,\ell,k}(z+2\pi R)=\tilde{\phi}_{a,m,\ell,k}(z)
\exp\left[-2\pi i\frac{nk}{N}\right]\qquad
(a=0, 1)\, ,
\ee
and this is gauge equivalent to $\tilde{\phi}_{a,m,\ell,k}(z)$ via
the transformation (\ref{gts}) with
\be
W_{k}=\exp\left(2\pi i\frac{nk}{N}\right)\, .
\ee
The degeneracy of each mass eigenvalue is $n$, which corresponds to
$\ell=1, \ldots , n$. 
This degeneracy is the same as the counterpart of continuum
theory~\cite{DS}.

In FIG.~\ref{fig1}, the mass-squared eigenvalues of the scalar field 
are exhibited against $fR$ for $n=N$. For general values of $n$, similar
dependence on $fR$ can be found. One notices that, in the limit of
$fR\rightarrow 0$, the mass-squared spectrum approaches the KK
spectrum with a circular dimensions, {\it i.e.} $m^2/R^2$ ($m$ :
integer).  This limit means that the discrete dimension becomes
degenerate.
Oppositely, in the limit of $fR\rightarrow \infty$,
the mass-squared spectrum behaves as $(2p+1)nf/(NR)$ ($p$ : integer).
This spectrum coincides with that of the continuum
theory~\cite{DS}.
One can find the boundary of the behavior of the mass level in
FIG.~\ref{fig1}. This lies on $M^2=4f^2$, which indicates how the mass
scale $f$ of the discrete compactification affects the mass spectrum.
Note that although the mass spectrum has both continuum and discrete
compactifications, that is not a mere sum of each spectrum.

Next, we consider the case that $n$ is not a divisor of $N$.
Let $g$ be the greatest common divisor of  $n$ and $N$.
In this case, not all $m$ is permitted but only $m=m'n/g$
$(m'=0, 1, 2, \ldots)$.
Then the eigenfunction is proportional to
\be
\tilde{\phi}_{a,m'n/g,\ell,k}(z)=\sum_{p'=1}^{N/g}
\phi_{a,m'n/g,\ell+p'g}(z)
\exp\left[2\pi i\frac{(\ell+p'g)k}{N}\right]\qquad
(a=0, 1)\, ,
\ee
and the degeneracy of each eigenvalue is given by $g$.
The restriction on the eigenstates looks very similar to the case
with orbifold compactification. 
Of course, the present case includes the previous case where $n$ is the
 divisor of $N$, as a special case.

\section{$SU(2)$ Yang-Mills field}
\label{sec:4}
In this section, we will briefly describe how the symmetry breaking
can occur in the Yang-Mills theory.
For simplicity, we consider a deconstucted $SU(2)$ Yang-Mills theory.
The action and the gauge symmetry on the fields are similar to the
$U(1)$ case in Sec.~\ref{sec:2}, provided that the fields 
$A_k^{\mu}$ and $\chi_k$ are matrix
valued and some trace operations are attached. Whereas the link field
$U_k$ is transformed by
$(SU(2))_k$ and
$(SU(2))_{k+1}$, we assume that the background link field takes the
following common $SU(2)$-valued matrix form:
\be
U_{k}=\frac{f}{\sqrt{2}}
\exp\left( i\frac{\tau_3}{2}\frac{nz}{NR}\right)
=\frac{f}{\sqrt{2}}
\left(
\begin{array}{cc}
e^{i\frac{nz}{2NR}} & 0\\
0 & e^{-i\frac{nz}{2NR}}
\end{array} \right)\, .
\ee
The possible $z$-dependent term for the Yang-Mills field comes from the
term
$\tr[(D^{\mu}U_{k})^{\dagger}D_{\mu}U_{k}]$ in the action.
We expand the field as
\be
\tilde{A}^{\mu}_{k}=\frac{1}{\sqrt{N}}\sum_{p=1}^NA^{\mu}_{p}
\exp\left[2\pi i\frac{pk}{N}\right]\, .
\ee
Moreover we can write $A^{\mu}_{p}=(A^{\mu}_{p})^a(\tau^a/2)$.
Among the three degrees of freedom in terms of $SU(2)$,
$(A^{\mu}_{p})^3$ has no $z$-dependent potential and
thus one vector field has the mass spectrum
\be
4f^2\sin^2\left(\frac{\pi p}{N}\right)+\frac{m^2}{R^2}\qquad\qquad p,
m~{\rm are~integer}\, .
\ee
At this point, the massive scalar fields which come from fluctuations
of $(\chi_p)^3$ in the link fields are absorbed in the vector field
except for one massless degree of freedom,
as in the case of deconstructed QED~\cite{HL1}.
 The rest two vector fields $(A^{\mu}_{p})^1$ and $(A^{\mu}_{p})^2$ have
the same spectrum as the scalar field in Sec.~\ref{sec:3}, because
the term $\tr[(D^{\mu}U_{k})^{\dagger}D_{\mu}U_{k}]$ reduces to
\bea
\tr[(D^{\mu}U_{k})^{\dagger}D_{\mu}U_{k}]&\rightarrow&
f^2\left|e^{\mp i\frac{nz}{2NR}}-e^{i\frac{2\pi
p}{N}}e^{\pm i\frac{nz}{2NR}}\right|^2\nn
&=&2f^2\left[1-\cos\left({\frac{2\pi
p}{N}\pm\frac{nz}{NR}}\right)\right]\, .
\eea
The $z$-component $(A^{z}_{p})^a$ becomes three scalar fields
which have the same mass spectrum as $(A^{\mu}_{p})^a$.
As a result, the massless sector in the $(D-1)$-dimensional theory
contains a
$U(1)$ gauge field and two scalar fields. 
\section{conclusion and discussion}
\label{sec:f}

In this paper, we have considered
a topologically non-trivial configuration for the link fields
in the deconstructed gauge theory with another compact
continuous dimension.
We have shown the bosonic spectrum in such a background fields.
We have also examined the possibility of non-Abelian symmetry breaking
by the background field by using a simple model.

A remarkable feature of the mass spectrum is as follows.
the spectrum of mass squared has a nearly equal intervals at low
levels, but another nearly equal intervals can be found in the spectrum
of mass at higher levels.

It is interesting to point out the difference between the mass of the
massive vector boson and that of the first excited states in the gauge
singlet; the former is of order $nf/(NR)$ or arbitrary small according
the choice of $fR$, $N$, and $n$, while the latter is the smaller one of
order either $1/R$ or $f/N$. In addition, we can naturally take another
sector independent of $f$. The KK excitation of such a sector has a
order of
$1/R$.
To summarize, we will be able to construct models with
seemingly additional mass scales from a few scales in a similar manner.

It should be studied the stability of the symmetry breaking vacuum.
Classically the vacuum with the non-trivial background field
has a positive finite energy density.
The Casimir-like energy may lower the value of the vacuum
energy~\cite{Hosotani}. For this purpose, we must consider also various
type of matter fields and their quantum effects.

In this paper, we have only treated the bosonic field.
Incidentally, the fermionic fields in the doubly latticized dimensions
with the non-trivial background fields have been studied since almost
three decades ago in the other context~\cite{Hof,GGHNR,KF}.
It is known that the eigenvalue equation become
a type of the Harper equation (or almost Mathieu equation)~\cite{KF,Wie}.
Therefore the doubly latticized space,
which is obtained if the $z$-direction in our present model is
discretized,
 can also be applied to the symmetry
breaking mechanism and will exhibit a new type of quantum effects
of fermionic and bosonic fields.
These subjects (on the Dirac fields and the doubly latticized extra
space) will be discussed elsewhere.

\begin{acknowledgments}
We would like to thank 
 K. Sakamoto 
for his valuable comments
and for the careful reading of the manuscript.
\end{acknowledgments}




\begin{thebibliography}{99}

\bibitem{KK} T.~Appelquist, A.~Chodos and P.~G.~O.~Freund,
{\it Modern Kaluza-Klein Theories}, Addison-Wesley, 1987.

\bibitem{ACG} N.~Arkani-Hamed, A.~G.~Cohen and H.~Georgi,
Phys. Rev. Lett. {\bf 86} (2001) 4757; 
Phys. Lett. {\bf B513} (2001) 232. 

C.~T.~Hill, S.~Pokorski and J.~Wang,
Phys. Rev. {\bf D64} (2001) 105005. 

\bibitem{HL1} C.~T.~Hill and A.~K.~Leibovich,
Phys. Rev. {\bf D66} (2002) 016006, 
{\tt hep-ph/0205057}.
\bibitem{HL2} C.~T.~Hill and A.~K.~Leibovich,
Phys. Rev. {\bf D66} (2002) 075010, 
{\tt hep-ph/0205237}.







\bibitem{RSS} S.~Randjbar-Daemi, A.~Salam and J.~Strathdee,
Nucl. Phys. {\bf B214} (1983) 491. 

\bibitem{DS} M.~J.~Duncan and G.~C.~Segr\`e,
Phys. Lett. {\bf B195} (1987) 36.

M.~J.~Duncan, G.~Segr\`e and J.~F.~Wheater,
Nucl. Phys. {\bf B308} (1988) 509.

K.~Zablocki,
J. Math. Phys. {\bf 29} (1988) 2653.

\bibitem{Manton} N.~S.~Manton, 
Nucl. Phys. {\bf B158} (1979) 141.

\bibitem{Hosotani} Y.~Hosotani,
Phys. Lett. {\bf B129} (1983) 193;
Phys. Rev. {\bf D29} (1984) 731.

\bibitem{KS} N.~Kan and K.~Shiraishi, ``Multi-graviton theory, a
latticized dimension and the cosmological constant'', {\tt
gr-qc/0212113}. 


\bibitem{GR} I.~S.~Gradstein and I.~M.~Ryshik,
{\it Tables of integrals, sums, series and products},
Academic Press, New York (1980).

M.~Abramowitz and I.~Stegun (eds.),
{\it Handbook of Mathematical Functions},
Dover, New York (1972).

\bibitem{Hof} D.~R.~Hofstadter,
Phys. Rev. {\bf B14} (1976) 2239.
\bibitem{GGHNR} L.~Giusti {\it et al.},
Phys. Rev. {\bf D65} (2003) 074506.
\bibitem{KF} H.~Kurokawa and T.~Fujiwara,
Phys. Rev. {\bf D67} (2003) 025015 and references there in.

\bibitem{Wie} A.~G.~Abranov, J.~C.~Talstra and P.~B.~Wiegmann,
Phys. Rev. Lett. {\bf 81} (1998) 2112;
Nucl. Phys. {\bf B575} (1998) 571.

P.~B.~Wiegmann, Prog. Theor. Phys. Suppl. {\bf 134} (1999) 171.











%
\end{thebibliography}
\end{document}